# Effect of the Bloch–Siegert Shift on the Frequency Responses of Rabi Oscillations in the Case of Nutation Resonance

A. P. Saiko[a] and G. G. Fedoruk[b]

[a] *Joint Institute of Solid State and Semiconductor Physics, Belarussian Academy of Sciences, Minsk, 220072 Belarus*
e-mail: saiko@ifttp.bas-net.by

[b] *Institute of Physics, University of Szczecin, 70-451 Szczecin, Poland*

The dynamics of a two-level spin system dressed by bichromatic radiation is studied under the conditions of double resonance when the frequency of one (microwave) field is equal to the Larmor frequency of the spin system and the frequency of the other (radio-frequency) field $\omega_{rf}$ is close to the Rabi frequency $\omega_1$ in a microwave field. It is shown theoretically that Rabi oscillations between dressed-spin states with the frequency $\varepsilon$ are accompanied by higher-frequency oscillations at frequencies $n\omega_{rf}$ and $n\omega_{rf} \pm \varepsilon$, where $n = 1, 2, \ldots$. The most intense among these are the signals corresponding to $n = 1$. The counter-rotating (antiresonance) components of the RF field give rise to a shift of the dressed-state energy, i.e., to a frequency shift similar to the Bloch–Siegert shift. In particular, this shift is manifested as the dependence of the Rabi-oscillation frequency $\varepsilon$ on the sign of the detuning $\omega_1 - \omega_{rf}$ from resonance. In the case of double resonance, the oscillation amplitude is asymmetric; i.e., the amplitude at the sum frequency $\omega_{rf} + \varepsilon$ increases, while the amplitude at the difference frequency $\omega_{rf} - \varepsilon$ decreases. The predicted effects are confirmed by observations of the nutation signals of the electron paramagnetic resonance (EPR) of $E'_1$ centers in quartz and should be taken into account to realize qubits with a low Rabi frequency in solids.

Resonant interaction between coherent radiation and quantum systems is most often described within the framework of the rotating-wave approximation. In this case, only one of two counter-rotating linearly polarized components of the electromagnetic field interacts efficiently with the quantum system, while the effect of the second component is negligible. When the electromagnetic field is strong, the counter-rotating (the so-called antiresonance) component becomes efficient and the rotating-wave approximation can no longer be used. Violation of the rotating-wave approximation results in the Bloch–Siegert effect [1]. This effect leading to a shift of the resonance frequency is negligible for optical transitions, but becomes significant for precision nuclear magnetic resonance (NMR) experiments [2]. The Bloch–Siegert effect can also give rise to variations important for the operation of quantum computers [3]. At the same time, because of the interference of the counter-rotating electromagnetic-field components, the Bloch–Siegert effect provides a wonderful opportunity to measure not only the amplitude, but also the absolute phase of a monochromatic electromagnetic field [4].

Interest has grown recently in studies of the interaction of two-level quantum systems with both continuous-wave [5, 6] and pulsed [7] bichromatic radiation formed by fields with strongly different frequencies.

Such an interest is explained by the importance and generality of the results for a wide range of physical objects, including, in particular, nuclear and electron spins in the NMR and EPR, double-well quantum dots, flux and charge qubits in superconducting systems, etc. Such systems are described with the use of the dressed-state approach by which the phenomenon of Rabi oscillations is analyzed. Note that the frequency of these oscillations depends on the efficiency of the interaction between the bichromatic radiation and quantum system, while their damping, on the coherence time of the quantum states. When quantum states dressed by the electromagnetic field are studied, the rotating-wave approximation is inevitably violated if the amplitude of the electromagnetic field exciting the transitions between the dressed states is comparable with the dressing-field amplitude. Hereafter, we consider electron two-level spin systems dressed by the bichromatic field formed by transverse microwave (MW) and longitudinal radio-frequency (RF) fields. It is known [7] that dressing by the electromagnetic field converts a two-level system into a dynamic multilevel system. Since the MW- and RF-field frequencies are drastically different, bichromatic radiation ensures the excitation of intense multiphoton transitions in the case of the EPR. The direct study of the dynamics of multiphoton transi-tions by the method of nonstationary nutation made it possible to measure the amplitudes of efficient fields between the spin states dressed by the RF field and to reveal the features of the effect similar to the Bloch–Siegert shift [8–11]. In the case of double resonance, where the frequency of the MW field is equal to the Larmor frequency of the spin system and the frequency of the RF field is close to the effective Rabi frequency (i.e., to the frequency of one-photon nutation) in the MW field, the RF field excites nutation (Rabi oscillation) between the spin states dressed by the MW field [12, 13]. In this case, the interaction of the spin system with its environment also changes drastically. An increase in the spin-coherence time under double-resonance conditions was recently revealed by the method of nonstationary nutations between the states of $E_1'$ centers in quartz and $P_1$ centers in a diamond dressed by the MW field [12–14]. Since the RF-field amplitude in these experiments is not much less than the MW-field amplitude, the rotating-field approximation for the RF field is violated and the effect of the counter-rotating RF-field component on the dynamics of the dressed-spin states becomes significant. However, the manifestation of a violation of the rotating-wave approximation was not analyzed in [12–14], although certain features that cannot be explained within the framework of the rotating-wave approximation were observed in the experimental dependences. It should be noted that an asymmetry of the observed-signal dependence on the RF-field frequency was discovered even in the first publication on the transitions between the dressed-spin states in the NMR (rotary saturation) [15]. The asymmetry of the Fourier spectrum of nutations on the dressed-electron spin states was also pointed out in [16], although the origin of these nutations remained unclear.

In this paper, using the example of nutations between the dressed-spin states of $E_1'$ centers in a crystalline quartz, we analyze theoretically and experimentally the features of the dynamics of dressed-spin states under EPR conditions in the case where the rotating-wave approximation is violated.

**Theory.** The Hamiltonian $H$ of the electron spin $S = 1/2$ in a MW field directed along the $x$ axis of the laboratory frame, as well as in a RF field and a static magnetic field that are directed along the $z$ axis, can be written as follows:

$$H = H_0 + H_x(t) + H_z(t). \quad (1)$$

Here, $H_0 = \omega_0 S^z$ is the Hamiltonian corresponding to the Zeeman energy of a spin in the magnetic field $B_0$, where $\omega_0 = \gamma B_0$ with $\gamma$ being the electron gyromagnetic ratio; $H_x(t) = 2\omega_1 \cos(\omega_{mw}t + \varphi)S^x$ and $H_z(t) = 2\omega_2 \cos(\omega_{rf}t + \psi)S^z$ are the Hamiltonians of the spin interaction with linearly polarized MW and RF fields, respectively; $B_1$ and $B_2$, $\omega_{mw}$ and $\omega_{rf}$, and $\varphi$ and $\psi$ are the amplitudes, frequencies, and phases of the MW and RF fields, respectively; $\omega_1 = \gamma B_1$ and $\omega_2 = \gamma B_2$ are the Rabi frequencies; and $S^{x,y,z}$ are the components of the spin operator.

First, we write Eq. (1) in the coordinate system rotating about the $z$ axis with a frequency of $\omega_{mw}$. Next, dropping rapidly oscillating ($e^{\pm i\omega_{rf}t}$) terms in the Hamiltonian by the Krylov–Bogoliubov–Mitropolsky averaging method [10], we obtain the following effective Hamiltonian describing $k$-photon ($k = 1, 2, 3, \ldots$) transitions between the dressed states:

$$H^{(k)} = (\Omega - k\omega_{rf} + \Delta_{BS})S^z + \frac{\omega_1 \omega_2 k(-1)^k}{2\Delta} J_k(a)(S^+ e^{-k\psi} + \text{H.c.}), \quad (2)$$

where $\Delta = \omega_0 - \omega_{mw}$, $\Omega = \sqrt{\Delta^2 + \omega_1^2}$, $a = 2\omega_2\Delta/\omega_{rf}\Omega$, and $\Delta_{BS} \approx \omega_2^2/4\omega_{rf}$ is the frequency shift similar to the Bloch–Siegert one. The absorption signal $v(t) = \frac{1}{2i}\text{Sp}\{\rho(t)(S^+ - S^-)\}$ is found from the Liouville equation $i\partial\rho/\partial t = [H^{(k)}, \rho]$ for the density matrix $\rho$. In the laboratory coordinate system, after averaging over the uniform distribution of the phase $\psi$ in the interval from 0 to $2\pi$, we obtain the following expression for $v(t)$ with $k = 1$:

$$v(t) \sim \frac{1}{2}\cos\theta\sin\xi J_{-1}(a)e^{-\Gamma t}\sin\varepsilon t + \frac{1}{8}\sin\theta J_0^2(a)e^{-\Gamma t}[2\sin^2\xi \sin\omega_{rf}t \quad (3)$$

$$+ (1-\cos\xi)^2\sin(\omega_{rf} - \varepsilon)t + (1+\cos\xi)^2\sin(\omega_{rf} + \varepsilon)t].$$

In this formula, the small terms oscillating with frequencies $n\omega_{rf}$ and $n\omega_{rf} \pm \varepsilon$, where $n = 2, 3, \ldots$, are disregarded;

$$\varepsilon = [(\Omega - \omega_{rf} + \Delta_{BS})^2 + \tilde{\omega}_2^2]^{1/2}, \quad (4)$$

where $\tilde{\omega}_2 = -(\omega_1\omega_{rf}/\Delta)J_1(a)$; $\sin\xi = \tilde{\omega}_2/\varepsilon$; $\cos\xi = (\Omega - \omega_{rf} + \Delta_{BS})/\varepsilon$; $\sin\theta = \omega_1/\Omega$; and $\cos\theta = \Delta/\Omega$. The frequency $\varepsilon$ given by Eq. (4) is the effective frequency of the Rabi oscillations between the spin states dressed by the MW field. The exponential decay of the signal with decay rate $\Gamma$ is introduced phenomenologically. To compare the theory with the experiment, formula (3) for the nutation signal should be averaged over the inhomogeneous frequency spread $\Delta$ (the parameters $a$, $\Omega$, $\Theta$, $\xi$, $\tilde{\omega}_2$, and $\varepsilon$ depend on $\Delta$) using the Gaussian weight function $g(\Delta) = (T_2^*/\sqrt{\pi})\exp(-\Delta^2 T_2^{*2})$:

$$\langle v(t) \rangle = \int_{-\infty}^{\infty} d\Delta g(\Delta) v(t), \quad (5)$$

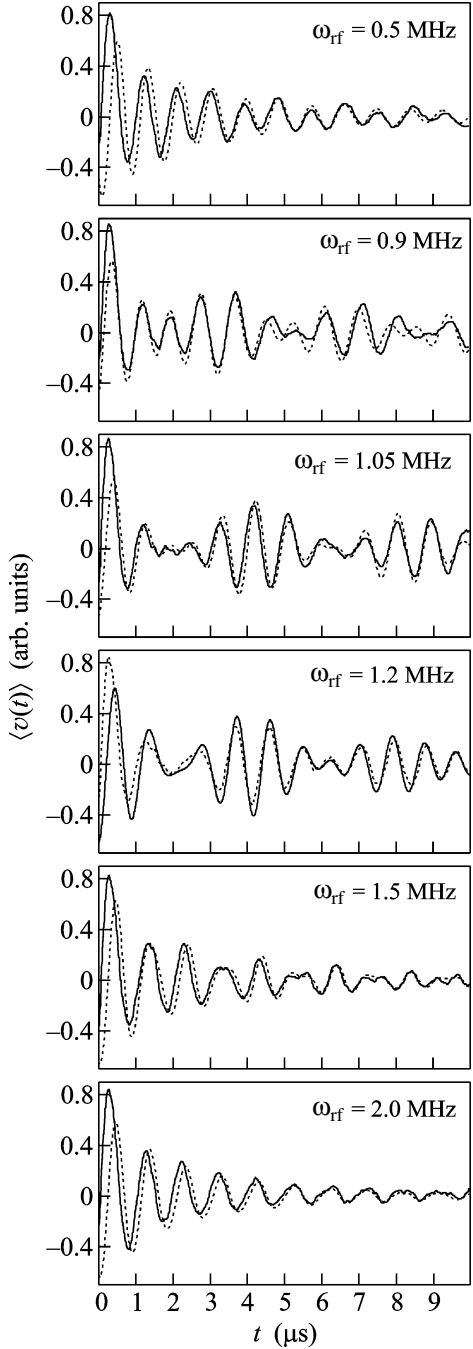

**Fig. 1.** Nutation signals of the EPR of $E'_1$ centers in quartz for various RF-field frequencies. Here, $\omega_{mw} = \omega_0$, $\omega_1/2\pi = 1.05$ MHz, and $\omega_2/2\pi = 0.24$ MHz. The solid and dashed lines correspond to the experiment and theory, respectively.

where $T_2^*$ is the time of the reversible phase relaxation.

It is seen in Eq. (3) that the Rabi oscillations at the frequency $\varepsilon$ between the dressed-spin states are accompanied by three higher-frequency oscillations at frequencies $\omega_{rf}$ and $\omega_{rf} \pm \varepsilon$. The Rabi-oscillation frequency depends on the Bloch–Siegert shift. For $\Delta = 0$ and $\Delta_{BS} \ll \omega_1, \omega_{rf}, \omega_2$, we have

$$\varepsilon \approx \varepsilon_0 \left[ 1 + \frac{(\omega_1 - \omega_{rf})\Delta_{BS} + \Delta_{BS}^2/2}{(\omega_1 - \omega_{rf})^2 + \omega_2^2} \right], \quad (6)$$

where $\varepsilon_0 = [(\omega_1 - \omega_{rf})^2 + \omega_2^2]^{1/2}$. If $|\omega_1 - \omega_{rf}| > \Delta_{BS}$, then the frequency $\varepsilon$ is a linear function of the shift similar to the Bloch–Siegert shift. In the case of nutation resonance where $\omega_{rf} = \omega_1$, this dependence is quadratic and, thus, much less pronounced, since $\Delta_{BS}^2$ is small. In the case of nutation resonance, the amplitude ratio of the signals at frequencies $\omega_{rf} - \varepsilon$ and $\omega_{rf} + \varepsilon$ is directly related to the Bloch–Siegert shift via the approximate formula

$$(1 - \cos\xi)^2/(1 + \cos\xi)^2 \approx 1 - 4\Delta_{BS}/\omega_2. \quad (7)$$

It follows from Eq. (7) that, in the presence of the Bloch–Siegert shift, the signal amplitude at the sum frequency is larger than the signal amplitude at the difference frequency and this asymmetry disappears in the absence of the Bloch–Siegert shift.

**Experimental results and discussion.** The nonstationary nutation of EPR signals was detected with the use of a 3-cm pulse EPR spectrometer [9]. The continuous-wave MW and RF fields were used. The stepwise onset of their resonant interaction with the spin system was realized by a pulse of a longitudinal magnetic field (the so-called Zeeman-pulse technique [9]).

The experiments were carried out at room temperature for the resonance static magnetic field ($\omega_{mw} = \omega_0$) ensuring the maximum absorption signal of nutations of the undressed spin system at the frequency $\omega_1$. The static magnetic field was parallel to the optical axis of the crystal. In this case, the EPR spectrum of the $E'_1$ centers comprises a single line of width $\Delta B_{pp} = 0.016$ mT. The duration, amplitude, and repetition period of the magnetic-field pulses were equal to 10 μs, $\Delta B = 0.12$ mT, and 1.25 ms, respectively. The Rabi frequencies $\omega_1$ and $\omega_2$ were calibrated with an accuracy of about 2% with the use of the frequencies of normal (in the absence of the RF field) and multiphoton (in a bichromatic field for $\omega_{rf} + \omega_{mw} = \omega_0$ [11]) nutations.

To improve the signal-to-noise ratio, the digital summation of the nutation signals obtained during each pulse was used. The phase of the RF field was not matched to the front edge of the magnetic-field pulse. Therefore, the observed signal is the result of the averaging of many (up to $10^3$) nutation signals with the uniform distribution of random phases of the RF field over the interval from 0 to $2\pi$. The same averaging was employed when deriving Eq. (3).

Figure 1 shows the nutation signals of the $E'_1$ centers in a crystalline quartz observed for various frequen-cies of the RF field and fixed MW- and RF-field amplitudes.

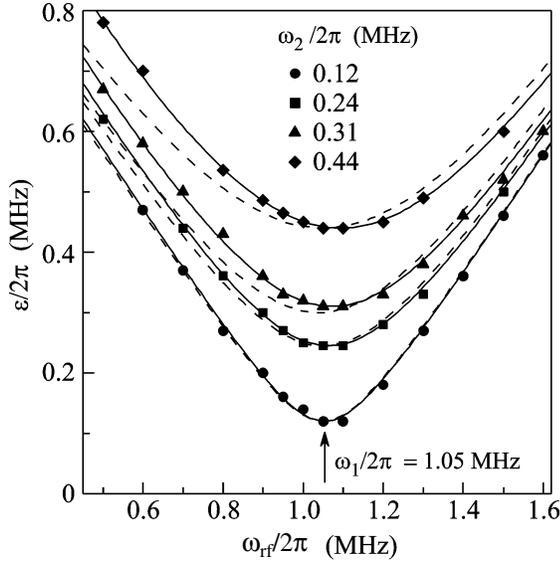

**Fig. 2.** Nutation frequency of the dressed-spin states versus the frequency of the RF field of various amplitudes for $\omega_{mw} = \omega_0$ and $\omega_1/2\pi = 1.05$ MHz. The solid and dashed lines correspond to the theory with and without the inclusion of the Bloch–Siegert effect, respectively.

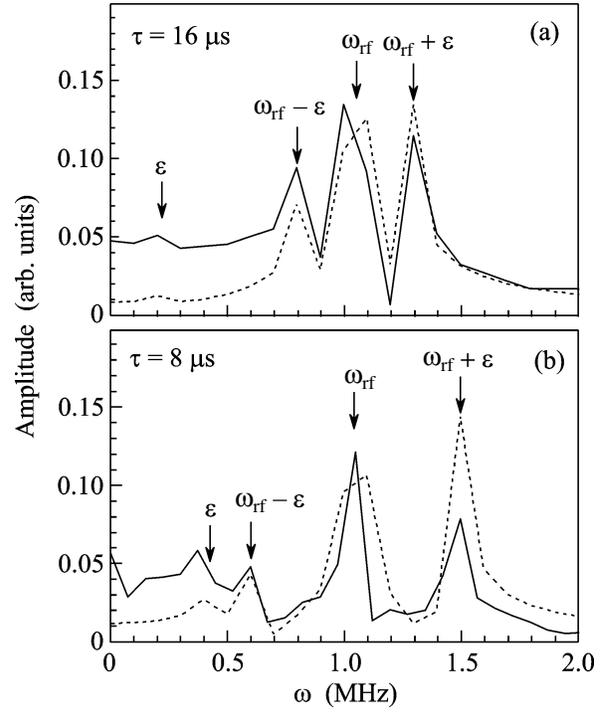

**Fig. 3.** Fourier spectra of the nutation signals observed for $\omega_1/2\pi = \omega_{rf}/2\pi = 1.05$ MHz and $\omega_2/2\pi =$ (a) 0.24 and (b) 0.44 MHz. The solid and dashed lines correspond to the experiment and theory, respectively. The arrows mark the oscillation frequencies calculated theoretically.

Note that, due to inhomogeneous broadening, the frequency of a one-photon nutation of the $E_1'$ centers excited in the absence of the RF field is independent of the detuning $\omega_{mw} - \omega_0$ from the nutation resonance and is equal to $\omega_1$. Nutation resonance in the presence of an RF field with a frequency approaching $\omega_1$ noticeably changes the observed nutation signal and enriches its spectral composition. The fitting of the observed signals by Eqs. (3)–(5) shown by the dashed lines in Fig. 1 demonstrates a good agreement between the theory and experiment and confirms that the nutation signal consists of a few components with frequencies $\varepsilon$, $\omega_{rf}$, and $\omega_{rf} \pm \varepsilon$.

It is seen in the signals shown in Fig. 1 and follows from Eq. (4) that the effective Rabi frequency $\varepsilon$ for the dressed-spin states depends on the detuning $|\omega_{rf} - \omega_1|$ from the nutation resonance. The value of $\varepsilon$ is minimum for $\omega_{rf} = \omega_1$ (exact resonance) and increases with an increase in the absolute value of the detuning. Moreover, it follows from Eq. (4) that $\varepsilon$ also depends on the frequency shift similar to the Bloch–Siegert shift. Since the RF-field amplitude is not negligibly small in comparison with the MW amplitude, the effect of the Bloch–Siegert effect under nutation-resonance conditions becomes observable. This shift is a nonresonance effect and, therefore, its manifestations are different for different signs of the detuning from resonance as seen from the comparison of the oscillograms obtained for $\omega_{rf}/2\pi = 0.9$ and 1.2 MHz ($|\omega_{rf} - \omega_1| = 0.15$ MHz).

The nutation frequencies of the dressed-spin states as functions of the frequency of the RF field with various amplitudes are presented in Fig. 2. The solid lines show the theoretical dependences of $\varepsilon$ on $\omega_{rf}$ with allowance for the Bloch–Siegert shift according to Eq. (4). For comparison, the same dependences disregarding the Bloch–Siegert shift are shown by the dashed lines. The data show that, owing to the Bloch–Siegert shift, the dependence of the frequency of the nutation of the dressed-spin states on the RF-field frequency is asymmetric with respect to $\omega_{rf} = \omega_1$ and this asymmetry becomes more pronounced as the RF-field amplitude increases.

Figure 3 shows the Fourier spectra of the nutation signals observed for two RF-field amplitudes under nutation-resonance conditions $|\omega_{rf} - \omega_1| = 0$. It is seen from the Fourier spectra of the observed signals and from Eq. (7) that, in the presence of the Bloch–Siegert shift, the amplitudes of these signals at frequencies $\omega_{rf} - \varepsilon$ and $\omega_{rf} + \varepsilon$ are not equal to each other even under nutation-resonance conditions. Note that a similar asymmetry of the side components was observed in the case of a three-photon nutation between the dressed nuclear spin states in NMR [17]. The reasons for its appearance were not analyzed in [17]. In our opinion, it should be attributed to the violation of the rotating-wave approximation. This can also explain the asym-metry of the nutation-resonance line profile in the case of its steady-state detection [15]. Moreover, it was shown theoretically in [18] that the asymmetry of the Autler–Townes doublet in the case of the optical observations of dressed nuclear spin states can also be attributed to the violation of the rotating-wave approximation.

Figure 1 demonstrates the unusual behavior of the relaxation of the dressed-spin states in the case of nutation resonance where $\omega_{rf} = \omega_1$. A comparison of the nutation decay rates shows that the dressed states interact with the environment much weaker than the undressed states. The approximation of the damping of the observed signals by an exponential with the decay rate $\Gamma$ shows that this quantity depends on the detuning $|\omega_{rf} - \omega_1|$ from the nutation resonance: $\Gamma$ is minimum for $\omega_{rf} = \omega_1$ (the decay time $\tau = 1/\Gamma = 16 \pm 2$ μs) and increases with a detuning up to the value $\tau = 2T_2$ typical of the normal one-photon nutation, where the spin–spin relaxation time is $T_2 = 3.6 \pm 0.4$ μs. For example, a fitting of the observed signals shown in Fig. 1 yields $\tau = 7, 14, 16, 14, 10,$ and $7$ μs for $\omega_{rf} = 0.5, 0.9, 1.05, 1.2, 1.5,$ and $2.0$ MHz, respectively. Note that it is the increase in the decay time of the nutation signal (narrowing of the line) that results in the dependence of the frequency $\varepsilon$ of the Rabi oscillations between the dressed-spin states on the detuning $|\omega_{rf} - \omega_1|$ from resonance. As was already mentioned above, the frequency of the normal nutation observed for this spin system in the absence of the RF field is equal to $\omega_1$ for any detuning $|\omega_{mw} - \omega_0|$.

Thus, it is proven theoretically and experimentally that the pulsed bichromatic excitation of a spin system under the conditions of double resonance (i.e., ordinary EPR, $\omega_{mw} = \omega_0$, and nutation resonance, $\omega_{rf} = \omega_1$) results in the appearance of Rabi oscillations with the frequency $\varepsilon$ between the spin states dressed by the MW field. These oscillations are accompanied by oscillations at frequencies $n\omega_{rf}$ and $n\omega_{rf} \pm \varepsilon$, where $n = 1, 2, \ldots$. The antiresonance effect of the RF field and, consequently, the violation of the rotating-wave approximation are manifested through an asymmetric dependence of the frequency $\varepsilon$ of the Rabi oscillation between the dressed-spin state on $\omega_{rf}$ and in the asymmetry of the amplitudes of signals at the frequencies $\omega_{rf} - \varepsilon$ and $\omega_{rf} + \varepsilon$ in the case of the exact nutation resonance ($\omega_{rf} = \omega_1$). These features are caused by a frequency shift similar to the Bloch–Siegert shift of the dressed-state energies. The mentioned effects should necessarily be taken into account for solid-state qubits with low Rabi frequencies (double-well quantum dots, flux and charge qubits in superconducting systems, etc.) in the case of their bichromatic excitation [6]. The observed increase in the coherence time of the dressed states under double-resonance conditions will be considered elsewhere.